\documentclass[prl,aps,twocolumn,showpacs,floatfix,superscriptaddress]{revtex4-1}
\usepackage{graphicx}
\usepackage{subfigure}
\usepackage{bm}
\usepackage{amsmath}
\usepackage{tabularx}
\usepackage{wasysym}
\usepackage{array}
\usepackage{epstopdf}
\usepackage{multirow}
\usepackage[version=3]{mhchem}
\usepackage{float}
\usepackage{color}
\usepackage{hyperref}
\usepackage{lipsum}
\usepackage{ulem}
\makeatletter
\newcommand{\colorcaption}[2][]{%
  \begingroup%
  \renewcommand{\@caption@fignum@sep}{ (Color online). }%
  \caption[#1]{#2}%
  \endgroup%
}
\makeatother

\usepackage{ulem}

\begin{document}

\title{Pre-melting \textbf{\textit{hcp}} to \textbf{\textit{bcc}} Transition in Beryllium}
\author{Y. Lu}
\affiliation{Beijing Computational Science Research Center, Beijing 100193, China}
\author{T. Sun}
\affiliation{ Key Laboratory of Computational Geodynamics, University of Chinese Academy of Sciences, Beijing 100049, China}
\author{Ping Zhang}
\affiliation{Institute of Applied Physics and Computational Mathematics, Beijing 100088, China}
\author{P. Zhang}
\affiliation{Department of Physics, University at Buffalo, State University of New York, Buffalo, New York 14260, USA}
\affiliation{Beijing Computational Science Research Center, Beijing 100193, China}
\author{D.-B. Zhang}
\thanks{Corresponding author}
\email[]{dbzhang@csrc.ac.cn}
\affiliation{Beijing Computational Science Research Center, Beijing 100193, China}
\author{R. M. Wentzcovitch}
\affiliation{Department of Chemical Engineering and Materials Science, and Minnesota Supercomputing Institute, University of Minnesota, Minneapolis, MN, 55455, USA}

\begin{abstract}

Beryllium (Be) is an important material with wide applications ranging from aerospace components to X-ray equipments.
Yet a precise understanding of its phase diagram remains elusive.
We have investigated the phase stability of Be using a recently developed hybrid free energy computation method that accounts for anharmonic effects by invoking phonon {quasiparticles}.
We find that the $hcp\rightarrow bcc$ transition occurs near the melting curve at $0<P<11$~GPa with a positive Clapeyron slope of $41\pm4$ K/GPa.
The $bcc$ phase exists in a narrow temperature range that shrinks with increasing pressure, explaining the difficulty in observing this phase experimentally. This work also demonstrates the validity of this theoretical framework based on phonon quasiparticle to study structural stability and phase transitions in strongly anharmonic materials.
\end{abstract}


\maketitle

Elemental solids usually undergo a series of phase transitions from ambient conditions to extreme conditions~\cite{element1,element2,element3}. Knowledge of their phase diagrams is a prerequisite for establishing their equations of state (EOS), a fundamental relation for determining thermodynamics properties and processes at high pressures and temperatures (PT). However, resolving phase boundaries is challenging for experiments given the uncertainties from several sources, especially at very high PT. Beryllium (Be) is a typical system whose phase diagram remains an open problem despite intense investigations. It assumes a hexagonal close-packed ($hcp$) structure at relatively low T~\cite{element1}.  Near the melting temperature $T_{M}$ ($\sim1,550$~K at 0 GPa), a competing phase with the body-centered cubic symmetry ($bcc$) seems to emerge~\cite{exp1,exp2,exp3}. However, not all experiments~\cite{hcp1,cexp1,cexp2,cexp3,cexp4,cexp5,cexp6} have observed this $bcc$ phase, causing confusion and controversies.
Be is important for both fundamental research~\cite{sci1,sci2,sci3,sci4} and practical applications.
Being a strong and light-weight metal, it has been widely used in a broad range of technological applications in harsh environments and extreme PT conditions, e.g., up to $T>$ 4,000 K and $P>$ 200 GPa~\cite{app1,app2,app3,app4,app5}.

The $bcc$ phase of Be was directly observed~\cite{exp1,exp2} only at $T~>1,500$~K around ambient pressure before melting.
Measurements of the temperature dependent resistance suggested that $bcc$ Be is a high pressure phase and the $hcp/bcc$ phase boundary between $0<P<6$~GPa has a negative Clapeyron slope ($-52\pm8$ K/GPa) ~\cite{exp3}. However, recent experiments have challenged this conclusion~\cite{cexp1,cexp2,cexp4,cexp5,cexp6}.
For example, it was reported that $bcc$ Be was not observed for $8<P<205$~GPa and $300<T<4,000$~K~\cite{cexp6}.
On the theory side, the study of Be's phase diagram using conventional methods encounters significant difficulties.
The lattice dynamics of $bcc$ Be is highly anharmonic, and the widely used quasi-harmonic approximation (QHA)
and Debye model are not able to capture such effect~\cite{qha1,qha2,qha3,debye1,debye2,debye3,debye4}.
For this reason, $bcc$ Be and the associated $hcp/bcc$ phase transition
remain poorly understood for $P <11$ GPa (density $<2.1g/cc$)
where $bcc$ Be is dynamically unstable at 0 K~\cite{qha2}.
At $P >11$ GPa, $bcc$ Be is dynamically stabilized by pressure and the QHA might, in principle, be applied.
However, the $hcp/bcc$ boundary~\cite{qha1,qha2,qha3} predicted by the QHA does not agree with experiments~\cite{cexp6},
suggesting that anharmonic effects still play an important role at higher pressures.

In this Letter, we report a new investigation of the phase stability of $bcc$ Be and the associated $hcp \rightarrow bcc$ phase
transition boundary up to 30 GPa and temperatures up to 2,000 K. We have used a recently developed hybrid approach ~\cite{tool1,tool2} that combines first-principles molecular dynamics (MD) and lattice dynamics calculations to address anharmonic effects in the free energy.
In this method, the concept of phonon quasiparticles offers a quantitative characterization of the effects of lattice anharmonicity~\cite{phonon1,phonon2}.
We show that Be exhibits pronounced anharmonic effects in both the $bcc$ and $hcp$ phases. Specifically, our results reveal the dynamical stabilization of $bcc$ Be with increasing T. The $bcc$ phase, however, is favorable only in a narrow temperature range near $T_M$, with $hcp\rightarrow bcc$ phase boundary having a positive Clapeyron slope of $41\pm4$ K/GPa.
The $bcc$ stability field shrinks with increasing pressure and eventually
disappears at around 11 GPa. This result seems to agree overall with early experiments~\cite{exp1,exp2} and differs from other $hcp/bcc$ phase boundaries (e.g., Mg \cite{renata3}), usually displaying a negative Clapeyron slope.

In the present approach, phonon quasiparticles are numerically characterized by the mode projected velocity autocorrelation function~\cite{tool1,tool2},
\begin{equation}
\langle{\bf V}(0)\cdot{\bf V}(t)\rangle_{q,s} = \lim_{t_0\rightarrow\infty}\frac{1}{t_0}\int_{0}^{t_0}{\bf V}_{q,s}^{\ast}(t')\cdot{\bf V}_{q,s}(t'+t)dt',
 \label{vaf}
\end{equation}
where ${\bf V}_{q,s}(t) = \sum_{i=1}^{N}\sqrt{M_i}{v}(t)\cdot \hat {\bf \epsilon}^{i}_{q,s}\exp(i{\bf q}\cdot {\bf R}_i)$
is the mode projected and mass weighted velocity for normal mode ($q,s$) with wave vector ${\bf q}$; $v_i(t) (i=1,...,N)$ is the atomic velocity produced by first-principles MD simulations with $N$ atoms, and $M_i$ and ${\bf R}_i$ are the atomic mass and coordinate
of atom $i$, respectively. $\hat{\bf \epsilon}^{i}_{q,s}(i=1,...,N)$ is the polarization vector of normal mode ($q,s$) calculated
using density functional perturbation theory (DFPT)~\cite{dfpt}.
For a well-defined phonon quasiparticle, the velocity autocorrelation function displays an oscillatory decaying behavior and its Fourier transform, i.e., the power spectrum,
   \begin{equation}
G_{q,s} = \int_{0}^{\infty}\langle{\bf V}(0)\cdot{\bf V}(t)\rangle_{q,s}\text {exp}(i\omega t)dt,
 \label{green}
\end{equation}
should have a Lorentzian-type line shape~\cite{tool1,tool2}.
The renormalized phonon frequency $\widetilde{\omega}_{q,s}$, and the linewidth, $\Gamma_{q,s}$ can then be obtained as discussed in more details in the Supplemental Materials.

The concept of phonon quasiparticle reduces the complex problem of interacting anharmonic phonons to an effective non-interacting system \cite{phonon1}, such that the conventional kinetic gas model and, to a great extent, the theory of harmonic phonons are still applicable. Moreover, since structural phase transition is triggered by lattice vibrations for many cases, insight into the transition mechanism can be also obtained by monitoring the variation of frequencies and line widths of phonon quasiparticles.

We used the Vienna $ab$ $initio$ simulation package (VASP)~\cite{vasp1,vasp2} to carry out first principles MD simulations on
$4\times4\times4$ supercells (128 atoms) of Be. We used the generalized gradient approximation of Perdew, Burke, and Ernzerhof~\cite{pbe} and the projector-augmented wave method~\cite{paw} with an associated plane-wave basis set energy cutoff of 350 eV. For metallic Be, the finite temperature Mermin functional ~\cite{mermin,mermin2} was used. Simulations were carried out at a series of volumes (V): $ 6.21 < V < 8.71 \text{\AA}^3$/atom for $bcc$ Be and $ 6.35 < V < 8.83 \text{\AA}^3$/atom for $hcp$ Be. For $hcp$ Be, proper aspect ratio ($c/a$) is adopted to obtain good hydrostatic conditions for specific volume and temperature. Temperatures ranging from 300 to 2,800 K are controlled through the Nos\'e dynamics~\cite{nose}. The considered volumes and temperatures result in a pressure range of  $0 < P < 30$ GPa.  For each volume and temperature, multiple independent MD runs (5 parallel replica) were performed to improve phase space sampling quality that also allow for evaluation of statistical uncertainties. Each MD run lasted 50 ps and used a time step of 1~fs. Harmonic phonon frequencies and normal modes were calculated using density functional perturbation theory (DFPT)~\cite{dfpt}.
\begin{figure}
\centering
\includegraphics[width=1\columnwidth]{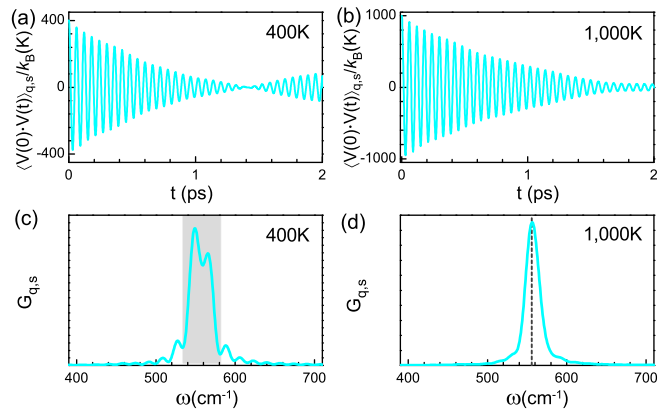}
\caption{ Mode projected velocity auto-correlation functions of the TA1 acoustic mode ($q,s$) at ${\bf q}=N$ with a harmonic frequency of 599~cm$^{-1}$ of $bcc$ Be at (a) 400~K and (b) 1,000 K, respectively. (c) and (d) show the corresponding power spectra. In (c), the shaded area between 535~cm$^{-1}$ and 580~cm$^{-1}$ covers two major peaks, indicating the breakdown of the phonon quasiparticle picture.
In (d), the vertical dashed line at 556~cm$^{-1}$ indicates the frequency of the well-defined phonon quasiparticle.}
\end{figure}

Before proceeding, we should clarify the general understanding of the $hcp\rightarrow bcc$ transition. The low T and low P $hcp$ structure relates to the $bcc$ structure through the zone center transverse optical (TO) mode and a macroscopic strain. This mode consists of opposite displacements of neighboring (0001) planes along $\langle1010\rangle$ and softens with increasing T. This is not necessarily a soft mode transition, but generally a first order transition with negative Clapeyron slope \cite{renata3}. The (0001) plane transforms into the (110) plane of the $bcc$ phase. This picture was validated by an early variable cell shape molecular dynamics study \cite{renata2}. The opposite $bcc\rightarrow hcp$ transition involves the lowest transverse acoustic mode (TA2) at ${\bf q} = [1/2, 1/2, 0]$, the $N$ point of the Brillouin zone,  marked by an open circle in Fig.~2(a). With this in mind we monitor closely the behavior of these modes with changing T.

We first investigate the behavior of phonon quasiparticles at different temperatures. For $bcc$ Be, phonon quasiparticles are not well defined at low T as found in other systems~\cite{phonon1}, but recovered at high T for unstable (soft) modes e.g., TA2. The analysis of this mode is shown in Fig.~S2 of the Supplementary material. It is more interesting to notice that at low T, phonon quasiparticles are not well defined even for certain stable modes with positive harmonic frequencies. Fig. 1 shows $\langle{\bf V}(0)\cdot{\bf V}(t)\rangle_{q,s}$ and the corresponding power spectra of the TA1 phonon mode at $N$ calculated at 400 K (Fig.~1(a)) and 1,000 K (Fig.~1(b)). This mode is marked by an open square in Fig.~2(a).
Although $\langle{\bf V}(0)\cdot{\bf V}(t)\rangle_{q,s}$ at $400$ K displays an oscillatory behavior, the  amplitude decay is non-monotonic (Fig.~1(a)). Consequently, the power spectrum has two major peaks within the shaded area as shown in Fig.~1(c).
This indicates that the frequency of this mode cannot be well constrained, or equivalently, the corresponding phonon quasiparticle is not well-defined. In contrast, at $1,000$~K $\langle{\bf V}(0)\cdot{\bf V}(t)\rangle_{q,s}$ exhibits a nicely decaying oscillatory behavior, Fig.~1(b). The corresponding power spectrum now has a well-defined Lorentzian line shape with a single and well defined peak, Fig.~1(d).
It is thus straightforward to identify the renormalized frequency of this mode as $556$~cm$^{-1}$. Similarly, all other quasiparticle mode frequencies sampled by the 4x4x4 supercell are equally well defined.
As previously indicated ~\cite{tool1,tool2}, these renormalized phonon frequencies plus the normal modes enable the calculation of the renormalized force constant matrix and complete phonon dispersions.
This quantitative characterization of phonon quasiparticles and renormalized phonon dispersion provide a solid foundation for studying thermal properties.

\begin{figure}
\centering
\includegraphics[width=1\columnwidth]{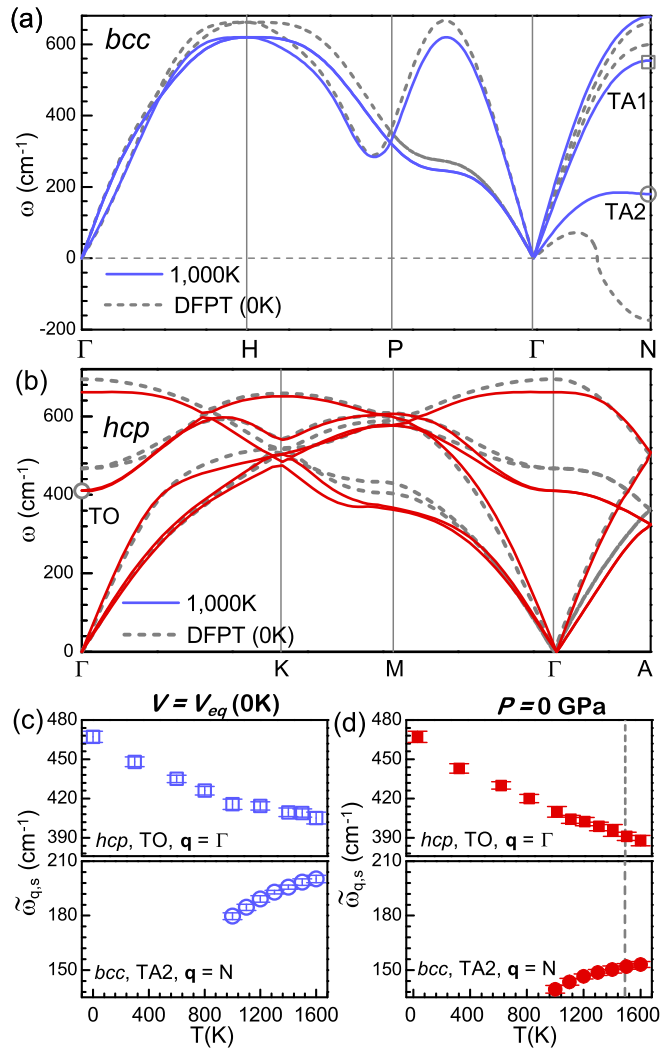}
\caption{ (a) Anharmonic phonon dispersion at 1,000~K (blue solid curves) and harmonic phonon dispersion calculated using DFPT (grey dashed curves) both at V = $7.81 \text{\AA}^3$/atom, the static $bcc$ Be equilibrium volume. The two transverse branches are labeled TA1 and TA2, respectively. (b) Anharmonic phonon dispersion calculated at $1,000$~K (blue solid curves) and harmonic phonon dispersion calculated using DFPT (grey dashed curves) both at $V = 7.89 \text{\AA}^3$/atom, the static $hcp$ Be equilibrium volume. (c) Temperature dependent frequency shifts of the TA2 ${\bf q}=$~N mode [open circle in (a)] and of the TO ${\bf q}=\Gamma$ mode [open circle in (b)] calculate at constant volume and (d) at constant zero pressure. The procedure to convert from constant volume to constant pressure was described in \cite{tool1}. The vertical dashed line in (d) indicates the $hcp/bcc$ transition temperature.}
\end{figure}

Figure~2(a) compares the anharmonic phonon dispersion of $bcc$ Be at 1,000~K with the harmonic phonon dispersion calculated using DFPT. Results are obtained at the static equilibrium volume of $bcc$ Be, $7.81 \text{\AA}^3$/atom.
There are noticeable differences between the anharmonic and harmonic phonon dispersions.
In particular, the unstable (soft) TA2 branch along the $\Gamma -N$ line stabilizes when high temperature anharmonic effects are accounted for.
This indicates that $bcc$ Be is stabilized by anharmonic effects.
To gain further insight into anharmonic effects, we analyze T-dependent phonon frequency shifts. Fig.~2(c) shows that the frequency of the $bcc$ zone edge phonon mode at ${\bf q}=N$ associated with the TA2 branch calculated at fixed volume varies non-linearly with T. Lowest order many-body perturbation theory (MBPT)~\cite{mbpt} predicts a linear frequency shift with T. Therefore, as expected, higher order anharmonic effects ignored in the perturbative treatment play an important role here.


The calculated anharmonic phonon dispersion over the whole Brillouin zone makes it possible to calculate the free energy in the thermodynamic limit ($N\rightarrow\infty$). Since $hcp$ Be is stable at low T for the entire pressure range of interest, the lattice thermal properties have been studied within the QHA~\cite{qha1,qha2,qha3} without further examination of the validity of the approximation. This naturally brings up a question: how important are anharmonic effects in the free energy in this seemingly stable structure? Fig.~2(b) compares the anharmonic phonon dispersion at $T = 1,000$~K and the harmonic phonon dispersion of $hcp$ Be calculated at a fixed volume of $7.89 \text{\AA}^3$/atom corresponding to zero static pressure.
The differences, although not alarming, are still significant in most of the Brillouin zone. A detailed analysis of individual phonon modes in the Supplementary material reveals that the frequency shifts with increasing T can be positive, negative or nearly zero, demonstrating the complexity of lattice anharmonic effects. What is important here is that the large frequency shifts in $hcp$ Be should be incorporated into the free energy calculation for more accurate evaluations of thermodynamic properties and phase boundaries.

The large frequency shifts not only reveal pronounced anharmonic effects but also shed light on the microscopic mechanism of this phase transition. As mentioned earlier, the $hcp$ and $bcc$ structures are related by a combination of phonon displacements and a macroscopic strain \cite{bain}. Together they provide a path for the $hcp\rightarrow bcc$ transition. The frequency of the zone center TO mode drops significantly from 467 to 405 cm$^{-1}$ when T increases from 0 to 1,600 K (see Fig.~2(c)). This observation is consistent with expectations based on the anticipated transformation mechanism ~\cite{renata1,renata2,renata3}. We note that although the frequency shift is very large at 1,600 K, the picture of phonon quasiparticle is still valid for the $hcp$ phase (see the Supplementary material for detailed analysis). As mentioned earlier, from the Brillouin zone-folding relation, the corresponding mode in $bcc$ Be is the zone edge TA2 mode at ${\bf q}=N$, whose property is shown in Fig.~2(c) and (d).



\begin{figure}
\includegraphics[width=1\columnwidth]{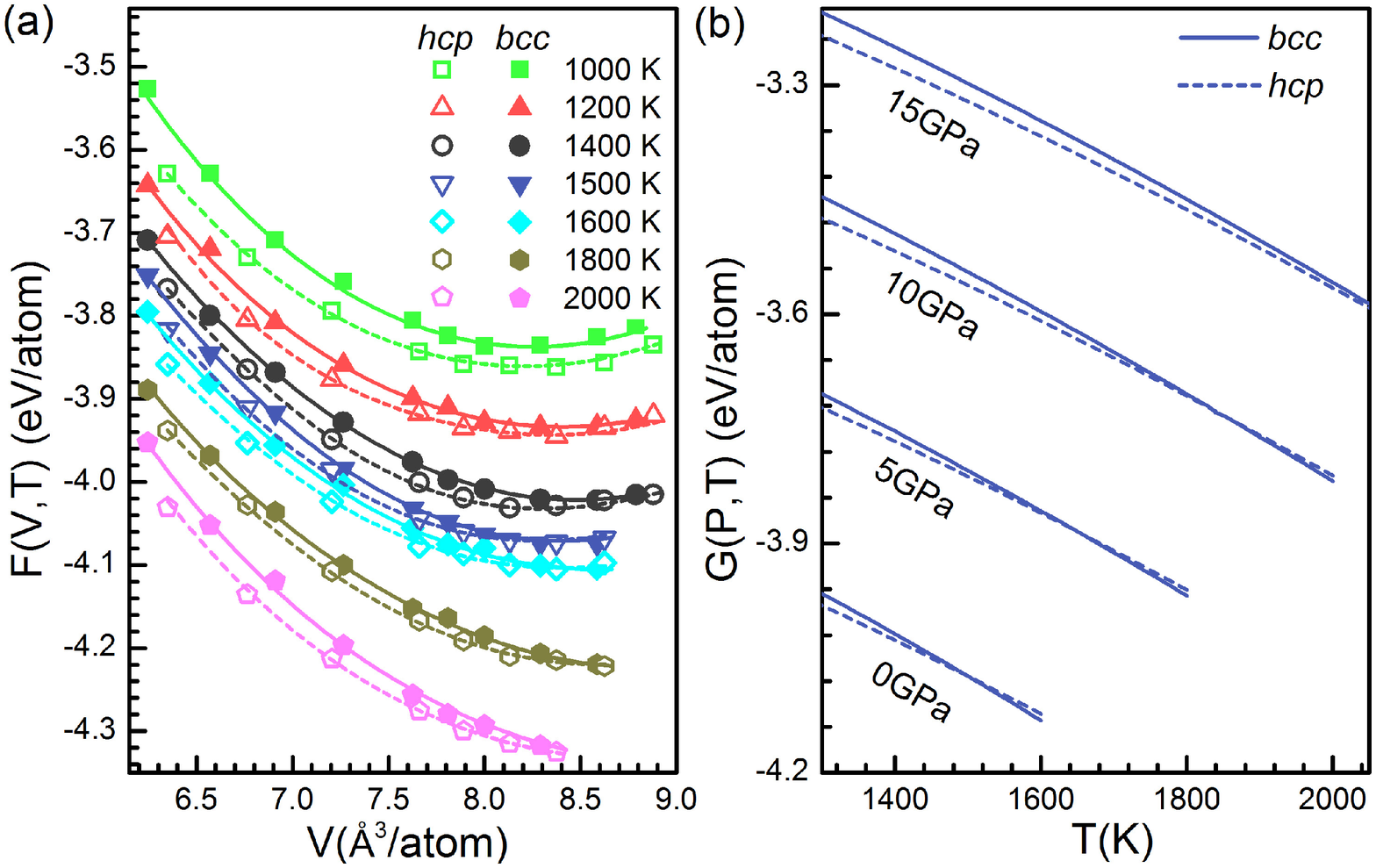}
\caption{ (a) Free energy $F(V,T)$ versus volumes of Be for $bcc$ (solid symbols) and $hcp$ (opened symbols) phases at different temperatures. (b) Free energy $G(P,T)$ versus temperature of Be for $bcc$ (solid lines) and $hcp$ (dashed lines) phases at different pressures. In (a) and (b), the error bars are too small to be visible.}
\end{figure}
\begin{figure}
\includegraphics[width=1\columnwidth]{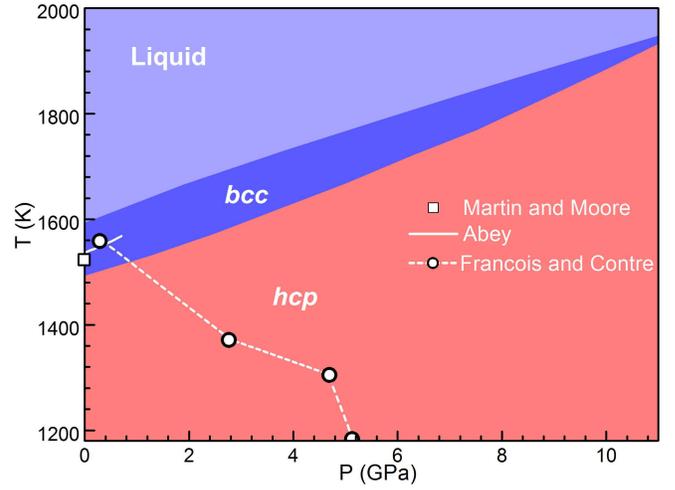}
\caption{ Phase diagram of Be. The melting line is adopted from Robert~\cite{qha2}. The experimental results for the $hcp/bcc$ phase boundary (solid line~\cite{exp2} and dashed line~\cite{exp3}) and the experimental point for $bcc$ Be (opened square)~\cite{exp1} are shown for comparison. }
\end{figure}

We now demonstrate that anharmonic effects are critical for obtaining this $hcp\rightarrow bcc$ phase boundary. When using anharmonic T-dependent phonon dispersions, the QHA free energy formula is no longer valid whereas the entropy formula is still applicable~\cite{helm}. Therefore, we first calculate the vibrational entropy~\cite{tool1,phonon1},
\begin{equation}
S_{vib}=k_{B}\sum_{\textbf{q}s}[(n_{\textbf{q}s}+1)\ln(n_{\textbf{q}s}+1)-n_{\textbf{q}s}\ln n_{\textbf{q}s}] ,
\end{equation}
with $n_{\textbf{q}s}=[\text{exp}(\hbar \tilde{\omega}_{\textbf{q},s}/k_{B}T)-1]^{-1}$,
and obtain the total free energy as:
\begin{equation}
F(V,T)= F(V,T_0)- \int_{T_0}^{T} S(T^{\prime}) dT^{\prime}.
\end{equation}
where $T_0$ is 1,000 K, $S$ is the total entropy including both vibrational and electronic contributions (See the Supplementary material for more details).  Our analysis of phonon quasiparticles demonstrates that they are well defined for both phases for $T\geq1,000$~K, therefore the choice of $T_0$. $F(V,T_0) = E(V,T_0)+ T_0S(V,T_0)$, where $E(V,T_0)$ is the internal energy obtained from the MD simulation. Fig.~3(a) displays the calculated free energies for both $bcc$ and $hcp$ phases. It is seen that at $T\geq1,200$ K, $V_{bcc} > V_{hcp}$, and consistently,  the common tangent to these curves starts to have negative slope, indicating a transition from $hcp$ to $bcc$ at positive pressure. We note that the volumetric variations of both phases also provide clues to understand the predicted $hcp/bcc$ transition at very high P (e.g., 400 GPa)~\cite{qha1,qha2,qha3}. More details of the variation of $V_{bcc}$ and  $V_{hcp}$ is shown in Fig.~S6 the Supplementary material.

It is more convenient to convert $F(V,T)$ into $G(P,T) = F(V,T) + P(V,T)V$ to obtain the phase boundary (see the Supplementary material for details). Fig.~3(b) displays $G(P,T)$ for both $bcc$ and $hcp$ phases. At each P, the intersection of $G_{bcc}$ and $G_{hcp}$ gives
the $hcp/bcc$ transition temperature. The resulting $hcp \rightarrow~bcc$ phase boundary shown in Fig.~4 together with the predicted/measured melting line, reveal several important aspects:
(i) $bcc$ Be is stable only when T approaches the melting point, in good agreement with direct experimental measurements.
For example, at ambient pressure where the melting temperature is $\sim1550$~K, $bcc$ Be was observed at $T>1500$~K~\cite{exp1,exp2}. (ii) The $hcp \rightarrow~bcc$ phase boundary has a positive Clapeyron slope of $41\pm4$ K/GPa, close to the experimental value ($43\pm7$ K/GPa) reported by Abey~\cite{exp2}.
Francois {\it et al.} reported a negative Clapeyron slope through an indirect measurement of the T-dependent resistivity~\cite{exp3}. Our results call for future experiments to clarify this issue.
(iii) The $hcp\rightarrow~bcc$ phase transition occurs only in a narrow pressure range of $0<P<11\pm2$~GPa. Our results are consistent with recent experiments that do not observe $bcc$ Be at high pressures~\cite{cexp1,cexp2,cexp3,cexp4,cexp5}. One report did not observe the $bcc$ phase at pressures as low as 8 GPa~\cite{cexp6} whereas our results suggest that the $bcc$ phase should exist up to 11 GPa. However, the predicted temperature range over which the $bcc$ phase is favorable is vanishingly narrow. Thus, it is possible that this narrow temperature range was missed in experiments or that our calculations overestimate the stability field of the $bcc$ phase by few GPa.

In summary, using the concept of phonon quasiparticle, we have investigated the controversial $hcp \rightarrow~bcc$ phase boundary of Be. We find that $bcc$ Be is stabilized at low pressures and high temperatures by anharmonic effects. For $hcp$ Be, anharmonic effects on phonon properties are
also significant. Using anharmonic phonon dispersions,
we evaluated the free energies of both phases and showed that the $bcc$ phase emerges as a pre-melting phenomenon at relatively low pressures.
Our results for the $hcp \rightarrow~bcc$ phase boundary are consistent with most experimental observations~\cite{exp1,exp2,cexp1,cexp2,cexp3,cexp4,cexp5,cexp6} in all important aspects. The temperature range over which the $bcc$ phase exists shrinks with increasing pressure and vanishes at a theoretical pressure of about 11 GPa. This narrow temperature range explains the difficulty in observing the $bcc$ phase experimentally.

This work was supported by NSFC under Grant Nos. U1530401, 41474069 and 11328401, and NSAF under Grant No. U1530258. RW was supported by the Hareaus visiting professorship award from the University of Frankfurt and by NSF grants EAR-1319368 and -1348066. PZ was supported by the NSF grant DMR-1506669.
Computations were performed at Beijing Computational Science Research Center and Minnesota Supercomputing Institute.


\begin{thebibliography}{99}

\bibitem{element1} J. F. Cannon, {\it J. Phys. Chem. Ref. Data}, {\bf 3}, 781 (1974).
\bibitem{element2} K. Persson, M. Ekman, V. Ozolin$\check{s}$, {\it Phys. Rev. B}, {\bf 61}, 11221 (2000).
\bibitem{element3}D. A. Young, {\it Phase Diagrams of the Elements}, University of California, 1991.

\bibitem{exp1} A. J. Martin and A. Moore, {\it J. Less-Common Met.} {\bf 1}, 85 (1959).
\bibitem{exp2} A. Abey, LLNL Report No. UCRL-53567, 1984 (unpublished).
\bibitem{exp3} M. Francois and M. Contre, in {\it Proceedings of the Conference internationale sur la metallurgie du beryllium}, Grenoble, (Presses Universitaires de France, Paris, 1965).


\bibitem{hcp1} W. J. Evans, M. J. Lipp, H. Cynn, C. S. Yoo, M. Somayazulu, D. H\"{a}usermann, G. Shen and V. Prakapenka, {\it Phys. Rev. B} {\bf 72}, 094113 (2005).
\bibitem{cexp1} L. C. Ming and M. H. Manghnani, {\it J. Phys. F: Met. Phys.} {\bf 14}, L1 (1984).
\bibitem{cexp2} V. Vijayakumar, B. K. Godwal, Y. K. Vohra, S. K. Sikka, and R. Chidambaram, {\it J. Phys. F: Met. Phys.} {\bf 14}, L65 (1984).
\bibitem{cexp3} A. R. Marder, {\it Science} {\bf 142}, 664 (1963).
\bibitem{cexp4} K. Nakano, Y. Akahama, and H. Kawamura, {\it J. Phys.: Condens. Matter.} {\bf 14}, 10569 (2002).
\bibitem{cexp5} N. Velisavljevic, G. N. Chesnut, Y. K. Vohra, S. T. Weir, V. Malba, and J. Akella, {\it Phys. Rev. B} {\bf 65}, 172107 (2002).
\bibitem{cexp6} A. Lazicki, A. Dewaele, P. Loubeyre, and M. Mezouar, {\it Phys. Rev. B} {\bf 86}, 174118 (2012).

\bibitem{sci1} S. H. Glenzer, G. Gregori, R.W. Lee, F. J. Rogers, S.W. Pollaine, and O. L. Landen, {\it Phys. Rev. Lett.} {\bf 90}, 175002 (2003).
\bibitem{sci2} S. H. Glenzer, O. L. Landen, P. Neumayer et al., {\it Phys. Rev. Lett.} {\bf 98}, 065002 (2007).
\bibitem{sci3} H. J. Lee, P. Neumayer, J. Castor et al., {\it Phys. Rev. Lett.} {\bf 102}, 115001 (2009).
\bibitem{sci4} I. Vobornik, J. Fujii, M. Hochstrasser et al., {\it Phys. Rev. Lett.} {\bf 99}, 166403 (2007)


\bibitem{app1} S. W. Haan {\it et al.}, {\it Phys. Plasmas} {\bf 18}, 051001 (2011).
\bibitem{app2} D. Swift, D. Paisley, and M. Knudon, {\it AIP Conf. Proc.} {\bf 706}, 119 (2003).
\bibitem{app3} K. L. Wilson, R. A Causey, W. L. Hsu, B. E. Mills, M. F. Smith, and J. B. Whitley, {\it J. Vac. Sci. Technol. A} {\bf 8}, 1750 (1990).
\bibitem{app4} D. S. Clark, S.W. Haan, and J. D. Salmonson, {\it Phys. Plasmas} {\bf 15}, 056305 (2008).
\bibitem{app5} D. C. Wilson, P. A. Bradley, N. M. Hoffman {\it et al.}, {\it Phys. Plasmas} {\bf 5}, 1953 (1998); J. L. Kline, S. A. Yi, A. N. Simakov  {\it et al.}, {\it Phys. Plasmas} {\bf 23}, 056310 (2015).

\bibitem{qha1} G. Robert and A. Sollier, {\it J. Phys. IV France} {\bf 134}, 257 (2006).
\bibitem{qha2} G. Robert, P. Legrand, and S. Bernard,  {\it Phys. Rev. B} {\bf 82}, 104118 (2010).
\bibitem{qha3} F. Luo, L.-C. Cai, X.-R. Chen, F.-Q. Jing, and D. Alf\'{e}, {\it J. Appl. Phys.} {\bf 111}, 053503 (2012).
\bibitem{debye1} L. X. Benedict, T. Ogitsu, A. Trave, C. J. Wu, P. A. Sterne, and E. Schwegler, {\it Phys. Rev. B} {\bf 79}, 064106 (2009).
\bibitem{debye2} G. V. Sin$'$ko and N. A. Smirnov, {\it Phys. Rev. B} {\bf 71}, 214108 (2005).
 \bibitem{debye3} K. K\'{a}das, L. Vitos, R. Ahuja, B. Johansson, and J. Koll\'{a}r, {\it Phys. Rev. B} {\bf 76}, 235109 (2007).
 \bibitem{debye4} K. K\'{a}das, L. Vitos, B. Johansson, and J. Koll\'{a}r, {\it Phys. Rev. B} {\bf 75}, 035132 (2007).

\bibitem{tool1} D.-B. Zhang, T. Sun, and R. M. Wentzcovitch, {\it Phys. Rev. Lett.} {\bf 112}, 058501 (2014).
\bibitem{tool2} T. Sun, D.-B. Zhang, and R. M. Wentzcovitch, {\it Phys. Rev. B} {\bf 89}, 094109 (2014).

\bibitem{phonon1} T. Sun, X. Shen, and P. B. Allen, {\it Phys. Rev. B} {\bf 82}, 224304 (2010).
\bibitem{phonon2} A. J. C. Ladd, B. Moran, and W. G. Hoover, {\it Phys. Rev. B} {34}, 5058 (1986).

\bibitem{renata3} J. D. Althof, P. B. Allen, R.. M. Wentzcovitch, and J. A. Moriarty, {\it Phys. Rev. B} {\bf 48}, 13253 (1993).

\bibitem{dfpt} S. Baroni, S. D. Gironcoli, A. D. Corso, and P. Giannozzi, {\it Rev. Mod. Phys.} {\bf 73}, 515 (2001); Giannozzi, et al., {\it J. Phys. Condens. Matter} {\bf 21}, 395502 (2009).


\bibitem{paw}P. E. Bl\"{o}chl, Phys. Rev. B {\bf 50}, 17953 (1994).

\bibitem{vasp1}
G. Kresse, J. Furthm$\ddot{\rm{o}}$ller, {\it Phys. Rev. B} {\bf 54}, 11169 (1999).
\bibitem{vasp2}G. Kresse and J. Hafner, {\it Phys. Rev. B} {\b 49}, 14251 (1994).


\bibitem{pbe}
J. P. Perdew, K. Burke, M. Ernzerhof, {\it Phys. Rev. B} {\bf 77}, 3865 (1996).
\bibitem{phononpy}A. Togo and I. Tanaka, {\it Scr. Mater.}, {\bf 108}, 1 (2015).
\bibitem{nose} S. Nos\'{e}, {\it J. Chem. Phys.} {\b 81}, 511 (1984); W.G. Hoover, {\it Phys. Rev. A} {\bf 31}, 1695 (1985).
\bibitem{mbpt} A. A.Maradudin and A. E. Fein, {\it Phys. Rev.} {\bf 128}, 2589 (1962).

\bibitem{bain}G. Grimvall, B. Magyari-Ko\"{p}e, Vidvuds Ozoli\c{n}\v{s}, and K. A. Persson, {\it Rev. Mod. Phys.} {\bf 84}, 945 (2012).

\bibitem{renata1} R. M. Wentzcovitch and M. L. Cohen, {\it Phys. Rev. B}, {\bf 37}, 5571 (1988).
\bibitem{renata2}  R. M. Wentzcovitch,  {\it Phys. Rev. B} {\bf 50}, 10358 (1994).

\bibitem{helm} D. C. Wallace, {\it Thermodynamics of Crystals} (Wiley, New York, 1972).
\bibitem{mermin}N. D. Mermin, {\it Phys. Rev.} {\bf 137}, A1441 (1965).
\bibitem{mermin2} R. M. Wentzcovitch, J. L. Martins, and P. B. Allen, {\it Phys. Rev. B} {\bf 45}, 11372 (1992).

\end{thebibliography}
\end{document}